\documentclass[aps,amsfonts,amsmath,prd,preprint,nofootinbib]{revtex4}
\newcommand{\beq}{\begin{equation}}
\newcommand{\eeq}{\end{equation}}

\usepackage{epsfig,bbm,cancel}
\usepackage[breaklinks=true]{hyperref}

\begin{document}

\title{Gravity of a noncanonical global monopole: conical topology and compactification}

\author{Ilham Prasetyo}
\email{ilham.prasetyo51@ui.ac.id}
\author{Handhika S. Ramadhan}
\email{hramad@ui.ac.id}
\affiliation{Departemen Fisika, FMIPA, Universitas Indonesia, Depok 16424, Indonesia. }

\def\changenote#1{\footnote{\bf #1}}

\begin{abstract}

We obtain solutions of Einstein's equations describing gravitational field outside a noncanonical global monopole with cosmological constant. In particular, we consider two models of k-monopoles: the Dirac-Born-Infeld (DBI) and the power-law types, and study their corresponding exterior gravitational fields. For each model we found two types of solutions. The first of which are global k-monopole black hole with conical global topology. These are generalizations of the Barriola-Vilenkin solution of global monopole. The appearance of noncanonical kinetic terms does not modify the critical symmetry-breaking scale, $\eta_{crit}$, but it does affect the corresponding horizon(s). The second type of solution is compactification, whose topology is a product of two $2$-dimensional spaces with constant curvatures; ${\mathcal Y}_4\rightarrow {\mathcal Z}_2\times S^2$, with ${\mathcal Y}, {\mathcal Z}$ can be de Sitter, Minkowski, or Anti-de Sitter, and $S^2$ is the $2$-sphere. We investigate all possible compactifications and show that the nonlinearity of kinetic terms opens up new channels which are otherwise non-existent. For $\Lambda=0$ four-dimensional geometry, we conjecture that these compactification channels are their (possible) non-static super-critical states, right before they undergo topological inflation.

\end{abstract}

\maketitle
\thispagestyle{empty}
\setcounter{page}{1}

\section{Introduction}

Monopole is a type of topological defects whose vacuum manifold $\mathcal{M}$ is non-contractible, $\pi_2(\mathcal{M})\neq\mathbb{I}$~\cite{Kibble:1976sj, vilenkinshellard}. Depending on the symmetry broken, monopole can be global or gauge. Gauge (magnetic) monopole has been predicted to exist as an elementary particle carrying magnetic charge in a certain models of grand unification~\cite{'tHooft:1974qc, Polyakov:1974ek}, and can be categorized as a topological soliton due to its regularity and energy finiteness~\cite{Manton:2004tk}. Global monopole, on the other hand, having Goldstone boson renders the energy to be linearly divergent; thus lacks the solitonic property in flat space.

When coupled to gravity, global monopole exhibits peculiar features. The most striking one, first studied by Barriola and Vilenkin~\cite{Barriola:1989hx}, is that it exerts no gravitational force on the matter around it (save from the tiny mass at the core\footnote{This tiny mass is later found to be negative, producing a small repulsive gravity~\cite{Harari:1990cz}.}) but the global geometry is not Euclidean; the space around monopole suffers from deficit solid angle $\Delta\equiv 8\pi G \eta^2$. The existence of deficit angle forces the spacetime around monopole to exist only when the symmetry-breaking scale $\eta$ is lower than its critical value $\eta_{crit}\equiv1/\sqrt{8\pi G}$. At $\eta=\eta_{crit}$, the deficit completely consumes the entire solid angle. It is suggested in Ref.~\cite{Olasagasti:2000gx} that the spacetime around critical global monopole may degenerate into a cylinder; the solid angle compactifies into a $2$-sphere. Only when $\Lambda\neq0$ does the radius of the $2$-sphere is determined by the theory; when $\Lambda=0$ the size becomes arbitrary and must be matched to an appropriate interior solution numerically. For higher-dimensional generalization of global monopole, however, numerical study does not seem to confirm this suggestion, as reported by Cho and Vilenkin~\cite{Cho:2003gn}. They observed that the radius (of the extra dimension) is a radially growing function. Instead, they found that this ``cigar geometry" solution can be obtained if they relax the requirement of staticity and let the monopole to inflate. Thus, compactification solution can be perceived as an inflating (non-static) super-critical solution written in some particular gauge. For the $4d$ case, it is shown numerically in Ref.~\cite{Liebling:1999bb} that regular solutions can still exist up to $\eta\lesssim\sqrt{3\over8\pi G}$, beyond which singularity develops. The failure to find static regular solutions above $\eta>\sqrt{3\over8\pi G}\approx0.345$ is interpreted as the appearance of topological inflation~\cite{Vilenkin:1994pv, Linde:1994hy}. Inflating global monopoles and its spacetime structure is studied in Ref.~\cite{Cho:1997rb}. However, no compactification solution has been found. It still remains unclear whether the super-critical global monopole can directly succumb to the formation of topological inflation or whether it develops spontaneous compactification as its intermediary fate.

From a completely different point of view, in field theory there has been a recent interest in the non-canonnical defects~\cite{Babichev:2006cy, Babichev:2007tn}. This was partly motivated by the search for inflaton within the framework of string theory~\cite{ArmendarizPicon:1999rj}. A subset of $k$-defects is the Dirac-Born-Infeld (DBI) defects~\cite{Babichev:2007tn, sarangi, Babichev:2008qv, pavlovski, ramadhan1, ramadhan2}, where kinetic term takes the form of Born-Infeld Lagrangian~\cite{Born:1934gh}. One advantage of DBI-form is its {\it naturalness}, that the theory does not possess any ad hoc higher-order kinetic terms which must be put by hand, and the fact that it can be thought as the stringy effect of going to the high-energy regime. 

In particular, global $k$-monopole has been studied in Ref.~\cite{Babichev:2006cy, Li:2005cp}, while its gravitational field is investigated in~\cite{Jin:2007fz, Liu:2009eh}. There, the authors obtained solutions for the metric and scalar fields numerically and showed that qualitatively the gravitational property of Barriola-Vilenkin (BV) monopole still holds; that the spacetime around the $k$-monopole still suffers from the solid deficit angle, and this angle depends not on the coupling constant of the $k$-terms but still on the symmetry-breaking scale $\eta$. The notable difference of these $k$-monopoles from their BV counterpart is that their mass can be negative or positive (which results in whether the gravitational field is repulsive or attractive), depending on the specific model of $k$-term considered. Despite their numerical results, the regime outside the monopole can be studied using the vacuum approximation, where the Higgs field is approximated to settle in the vacuum manifold, $|\phi|\approx\eta$. In this approximation the analytical solutions can be found. 

Our aim in the present work is therefore twofolds. First, we look for analytical solutions of gravitational field with the same metric ansatz as the BV monopole. In particular we study the case when a $k$-monopole is swallowed by a black hole. This results in the Reissner-Nordstrom-type of black hole, only that the charge is now provided by the scalar field. Second, we look for solutions of different type of ansatz, that is the one where the $4d$ spacetime can be written as a product of two $2$-spaces of constant curvature. These results can be thought of spontaneous compactifications of four-dimensional maximally-symmetric spaces into two-dimensional using the global $k$-monopole. We show that our result generalizes that of Olasagasti and Vilenkin~\cite{Olasagasti:2000gx} in four-dimensional spacetime.

This paper is organized as follows. The next Section is devoted to the black hole solutions of Einstein's equations for $k$-monopoles with $\Lambda$. Throughout this work we focus only on two types of $k$-terms: the DBI and the power-law types. In Section III we consider a compactification ansatz for the metric, where the radius is held fixed. Here we list the possible compactification channels from the $4d$ down to $2d\times 2d$. Finally, our conclusions are summarized in Section IV.

\section{Black hole solutions}
\subsection{Gravitational field of Dirac-Born-Infeld monopole}

We begin with the action~\cite{Liu:2009eh, Li:2002ku, Bertrand:2003yq}
\begin{equation}
\mathcal{S}=\int d^4x\sqrt{-g}\left({\mathcal{R}-2\Lambda\over 16\pi G}+\mathcal{L}_{DBI}\right),
\end{equation}
whose Lagrangian is given by,
\begin{equation}
\mathcal{L}_{\text{DBI}}=
\beta^2 \left(
1- \sqrt{1-\frac{\partial^\mu \phi^a \partial_\mu \phi^a}{\beta^2}}
\right) - \frac{\lambda}{4} (\phi^a \phi^a - \eta^2)^2,
\label{eq005}
\end{equation}
here $\beta$ parametrizes the nonlinearity of our theory; $\beta\rightarrow\infty$ reduces the Lagrangian above into the ordinary global monopole Lagrangian.  

We use the following ansatz for the metric
\begin{equation}
ds^2=A(r)^2 dt^2-B(r)^2 dr^2-r^2\left(d\theta^2+\sin^2\theta\ d\phi^2\right),
\end{equation}
and for the scalar field
\begin{equation}
\phi^a=\eta f(r) \frac{x^a}{r}.
\end{equation}
From the metric, we get the components of Ricci tensor
\begin{eqnarray}
&&R^t_t=\frac{1}{B^2}\left(\frac{A''}{A}-\frac{A'B'}{AB}+\frac{2A'}{rA}\right),\\
&&R^r_r=\frac{1}{B^2}\left(\frac{A''}{A}-\frac{A'B'}{AB}-\frac{2B'}{rB}\right),\\
&&R^\theta_\theta=\frac{1}{B^2}\left(\frac{1}{r^2}+\frac{A'}{rA}-\frac{B'}{rB}\right)-\frac{1}{r^2}=R^\phi_\phi.
\end{eqnarray}
The prime denotes derivative with respect to $r$. From the Lagrangian, we get energy-momentum tensor components
\begin{eqnarray}
&&T^t_t=-\beta^2 \left(1-{1\over\gamma}\right)+\frac{\lambda\eta^4}{4}(f^2-1)^2+\frac{\Lambda}{8 \pi G},\label{eq0010}\\
&&T^r_r=T^t_t-{\eta^2\gamma f'^2\over B^2},\\
&&T^\theta_\theta=T^t_t-{\eta^2 f^2\over r^2}=T^\phi_\phi,
\end{eqnarray}
along with the equation for the scalar field 
\begin{eqnarray}
{1\over ABr^2} \left[{Ar^2\gamma f'\over B}\right]'-{2\gamma f\over r^2}-\lambda\eta^2f(f^2-1)=0,
\end{eqnarray}
where $\gamma\equiv{1\over \sqrt{1+\left({\eta^2\over\beta^2}\right)\left({f'^2\over B^2}+{2f^2\over r^2}\right)}}$.

We are interested only in exterior solution, $f\approx 1$, which makes
\begin{eqnarray}
&&T^t_t=T^r_r=-\beta^2\left(1-\sqrt{1+{2\eta^2\over\beta^2 r^2}}\right)+{\Lambda\over 8 \pi G},\\
&&T^\theta_\theta=T^\phi_\phi=T^t_t
-{\eta^2/r^2\over\sqrt{1+{2\eta^2\over\beta^2 r^2}}}.
\end{eqnarray}
Because $T^t_t=T^r_r$, $R^t_t=R^r_r$, then we get $B=A^{-1}$. From Einstein equation $R^t_t-\frac{1}{2}\delta^t_t R = 8\pi G T^t_t$,
\begin{equation}
{1\over r^2}\left[1-(rB^{-2})'\right]=-8\pi G \beta^2\left(1-\sqrt{1+{2\eta^2\over\beta^2 r^2}}\right)+\Lambda,
\end{equation}
which can be integrated to yield
\begin{equation}
B^{-2}=1-\frac{2GM}{r}+{(8\pi G\beta^3-\Lambda)\over 3}r^2-\frac{8}{3}\pi G\beta^2 r^2 \left(1+{2\eta^2\over r^2\beta^2}\right)^{3/2},
\label{eq001}
\end{equation}
with $M$ a constant of integration. 
This metric is not asymptotically flat, even in the absence of $\Lambda$. Instead, as in the case of ordinary global monopole~\cite{Barriola:1989hx} the metric develops conical global topology, as we shall see later. One way to see this is by evaluating the scalar curvature (for $\Lambda=0$), which is non-zero
\begin{equation}
R={16\pi G\left(-3\eta^2+2\beta^2 r^2\left(-1+\beta\sqrt{1+{2\eta^2\over\beta^2 r^2}}\right)\right)\over r^2\sqrt{1+{2\eta^2\over\beta^2 r^2}}}.
\end{equation}
It is amusing to note that in the limit $r\rightarrow\infty$ the scalar is constant, $R=32\pi G\beta^2\left(\beta-1\right)$.

Since we are looking for the exterior solution, our metric is valid only for $r>\delta$, where $\delta$ is the physical size of the monopole. This is the regime where the scalar field relaxes to its vacuum value ($f\approx1$). The size $\delta$ can be roughly estimated to be, for example see Ref.~\cite{Liu:2009eh}, as follows. From the Einstein's equations we get (after assuming $f(0<r<\delta)\approx 0$ and $f(r>\delta)\approx 1$)
\begin{equation}
B^{-2}=1-\frac{8\pi G}{r}\int\limits^{\delta}_{0} r^2 T^t_t(f\approx 0) dr -\frac{8\pi G}{r}\int\limits^{r}_{\delta} r^2 T^t_t(f\approx 1) dr,
\end{equation}
and assuming the metric solution to be
\begin{equation}
B^{-2}\equiv 1-8\pi G \eta^2 -\frac{2G\eta}{r} \mathcal{M}(r),
\end{equation}
with $\mathcal{M}(r)$ the mass of the core which is not the same as $M$, we will get 
\begin{equation}
\mathcal{M}(r)= -4\pi\eta r+\frac{\pi \lambda \eta^3 \delta^3}{3} -\frac{4\pi\beta^2(r^2-\delta^2)}{3\eta} +\frac{4\pi\beta^2}{3\eta}\left[
\left(r^2+\frac{2\eta^2}{\beta^2}\right)^{3/2}-\left(\delta^2+\frac{2\eta^2}{\beta^2}\right)^{3/2}
\right]+\frac{\Lambda r^3}{6 G\eta},
\end{equation}
and calculating $d\mathcal{M}(r)/d\delta=0$ will give us
\begin{equation}
\label{dbisize}
\delta=\frac{1}{\sqrt{\frac{\lambda^2\eta^6}{32\beta^2}+\frac{\lambda\eta^2}{4}}}.
\end{equation} 
Thus, in the weak-coupling regime, $\beta^2\gg 1$ the size reduces to that of~\cite{Barriola:1989hx}; $\delta\sim\lambda^{-1/2}\eta^{-1}$. On the other hand for the strong-coupling regime, $0<\beta^2\ll 1$, the size gets thinner, as in~\cite{Liu:2009eh}; $\delta\sim\beta\lambda^{-1}\eta^{-2}$. The nonlinearity of DBI theory enables the global monopole to relax faster to its vacuum value.

To get a better understanding of what Eq.~\eqref{eq001} describes we may expand the metric with respect to $r$. Keeping only the terms up to $O\left(r^{-2}\right)$ the metric can be cast into
\begin{eqnarray}
\label{dbiapprox}
ds^2&\simeq&\left(1-\Delta-{2GM\over r}-{4\pi G\beta^{-2}\eta^4\over r^2}-{\Lambda\over 3}r^2\right)dt^2-{dr^2\over\left(1-\Delta-{2GM\over r}-{4\pi G\beta^{-2}\eta^4\over r^2}-{\Lambda\over 3}r^2\right)}\nonumber\\
&&+r^2 d\Omega^2_2,
\end{eqnarray}
where $\Delta\equiv 8\pi G\eta^2$. This resembles a "charged" black hole\footnote{This charge is not of Reissner-Nordstrom type, since the term proportional to $1/r^2$ has minus sign. This manifests in the $\Lambda=0$-case where the black hole has only one horizon, as we shall see below.}, only that it develops deficit angle which results in the conical global topology. This deficit angle can be best seen by the following rescaling\footnote{This rescaling can in fact be applied to a more general higher-dimensional black hole solutions~\cite{Tangherlini:1963bw, Gregory:1995qh}. Suppose the metric of a spherically-symmetric $p$-brane in $N$-dimensions ($N=D+p$),
\begin{equation}
\label{metricfoot}
ds^2=A(r) g_{\mu\nu}dx^{\mu}dx^{\nu}-{dr^2\over A(r)}-r^2d\Omega^2_{D-2},
\end{equation}
has a solution
\begin{equation}
A(r)=1-\Delta-\sum^{\infty}_{i=1}\alpha_i r^i-\sum^{\infty}_{j=1}\sigma_j r^{-j},
\end{equation}
where $\alpha_i$ and $\sigma_j$ are constant coefficients. Under the rescaling $x^{\mu}\rightarrow\left(1-\Delta\right)^{1/2}x^{\mu}$ and $r\rightarrow{r\over\left(1-\Delta\right)^{1/2}}$, the metric\eqref{metricfoot} can be cast as a black brane with conical topology,
\begin{equation}
ds^2=A(r) g_{\mu\nu}dx^{\mu}dx^{\nu}-{dr^2\over A(r)}-\left(1-\Delta\right)r^2d\Omega^2_{D-2},
\end{equation}
with $A(r)\rightarrow 1-\sum^{\infty}_{i=1}\alpha_i r^i-\sum^{\infty}_{j=1}\sigma_j r^{-j}$, provided $\alpha_i\rightarrow\alpha_i\left(1-\Delta\right)^{i-2\over 2}$ and $\sigma_j\rightarrow\sigma_j\left(1-\Delta\right)^{-(j+2)\over 2}$.
}(for example, see~\cite{Marunovic:2013eka}):
\begin{equation}
t\rightarrow\left(1-\Delta\right)^{1/2}t,\ \ \ r\rightarrow {r\over\left(1-\Delta\right)^{1/2}}.
\end{equation}
This renders the metric rescaled as follows
\begin{eqnarray}
\label{dbiapproxlambda}
ds^2&=&\left(1-{2GM\over r}-{4\pi G\beta^{-2}\eta^4\over r^2}-{\Lambda\over 3}r^2\right)dt^2-{dr^2\over\left(1-{2GM\over r}-{4\pi G\beta^{-2}\eta^4\over r^2}-{\Lambda\over 3}r^2\right)}\nonumber\\
&&+\left(1-\Delta\right)r^2 d\Omega^2_2,
\end{eqnarray}
provided we simultaneously rescale $G\rightarrow{G\over\left(1-\Delta\right)^2}$ and $M\rightarrow\left(1-\Delta\right)^{1/2}M$. This metric, under the condition of
\begin{equation}
\label{bhcon}
GM\gg\delta
\end{equation}
describes a de Sitter black hole carrying global monopole charge. If this condition is not met then the solution describes the spacetime outside a regular DBI monopole, as shown by~\cite{Barriola:1989hx} for the case of ordinary global monopole. 

As suggested in~\cite{Barriola:1989hx} this metric describes a DBI global monopole being swallowed by a de Sitter black hole. The last term clearly shows that it has deficit solid angle; the area of the sphere with radius $r$ now becomes $4\pi\left(1-\Delta\right)r^2$. We can define the critical symmetry-breaking scale $\eta_{crit}\equiv(8\pi G)^{-1/2}$. The solution~\eqref{dbiapproxlambda} is valid only for $\eta<\eta_{crit}$, beyond which the deficit angle eats up the whole area of the sphere. 

For the case of $\Lambda=0$ this black hole seems to have two roots. They are 
\begin{equation}
r_\pm=GM\left(1\pm \sqrt{1+\frac{4\pi\eta^4}{M^2 G\beta^2}}\right),
\end{equation}
But do not be deceived. The inner one ($r_{-}$) cannot exist; for any values of $\eta$ and $\beta$, $r_-<0$. Thus only one root is physical. This confirms our claim that this black hole is not of the Reissner-Nordstrom type. To be a black hole at all, we must also check whether $r_+$ is really a horizon; whether it lies outside or inside the monopole. The black hole condition thus requires $\delta<r_+$. From condition~\eqref{bhcon} we obtain 
\begin{equation}
1+\sqrt{1+{4\pi\eta^4\over M^2 G\beta^2}}\gg0.
\end{equation}
This can be satisfied for any value of $\beta>0$, which means that the horizon is bound to exist.

We should also check whether the condition~\eqref{bhcon} can be satisfied in our case. With $\delta$ given by ~\eqref{dbisize} we can read off
\begin{equation}
{\lambda^2\eta^6\over 32\beta^2}+{\lambda\eta^2\over 4}> {1\over G^2 M^2}.
\end{equation}
This results in
\begin{equation}
\beta=
\begin{cases}
<{1\over 2\sqrt{2}}\sqrt{G^2M^2\eta^6\lambda^2\over 4-G^2M^2\eta^2\lambda},& \text{if~} 0<\eta<{2\over GM\lambda^{1/2}},\\
>0,& \text{if~} \eta>{2\over GM\lambda^{1/2}}.
\end{cases}
\end{equation}
For a reasonable value of $\lambda$ we must have $M$ $\gtrsim m_P$. This should not trouble us since the original Barriola-Vilenkin global-monopole black hole~\eqref{bhcon} also requires $M\gtrsim m_P$~\cite{Barriola:1989hx}. Thus the existence of DBI-monopole black hole endowed with global charge is fairly generic in our solution.

Let us now consider $\Lambda\neq 0$. Metric~\eqref{dbiapprox} has an additional horizon, the cosmological horizon, given by the roots of 
\begin{equation}
{\Lambda r^4\over 3}-r^2+2 G M r+{4\pi G\eta^4\over\beta^2}=0.
\end{equation}
This is a quartic equation whose actual roots are complicated, and we avoid writing them explicitly. Instead one can look at the almost-purely global monopole-de Sitter case ($\Lambda\gg M^{-2}$). Here we can find two horizons corresponding to the roots of
\begin{equation}
\label{approxroot}
\rho^2-{3\over \Lambda}\rho-{12\pi G\eta^4\over\beta^2\Lambda}\simeq0,
\end{equation}
where $\rho\equiv r^2$. They are given by
\begin{equation}
r_{\pm}\simeq\sqrt{3\over 2\Lambda}\sqrt{1\pm\sqrt{1-{16\pi G\eta^4\Lambda\over 3\beta^2}}},
\end{equation}
the global-monopole and cosmological horizons, respectively. For $\Lambda>0$ the two horizons can coalesce when $\eta\equiv\eta_{ext}=\left({3\beta^2\over 16\pi G\Lambda}\right)^{1/4}$, or equivalently when $\Lambda\equiv\Lambda_{ext}={3\beta^2\over 16 G\pi\eta^4}$. When $\eta$ or $\Lambda$ exceed their extremal values then we are left with naked singularity. For $\Lambda<0$ the cosmological horizon becomes complex and we are left only with the inner horizon.

\subsection{Gravitational field of k-monopole}

In this section, we show an explicit example which begins with Lagrangian
\begin{equation}
\mathcal{L}_K=K(X)-{\lambda\over 4}
(\phi^a\phi^a-\eta^2)^2.
\end{equation}
with the scalar field and the metric tensor the same as the previous section, thus
\begin{equation}
X=-{1\over 2}\partial^\mu \phi^a \partial_\mu \phi^a=\frac{\eta^2}{2}\left({f'^2\over B^2}+{2f^2\over r^2}
\right).\label{eq009} 
\end{equation}

We consider 
\begin{equation}
K(X)=-X-\beta^{-2} X^2.\label{eq008}
\end{equation}
with $\beta^{2}>0$. The weak-field limit $K(X)\rightarrow -X$ is achieved when $\beta^{2} \rightarrow \infty$. The components of energy-momentum tensor are
\begin{eqnarray}
&&T^t_t=X+\beta^{-2} X^2+{\lambda\eta^4\over 4}(f^2-1)^2+{\Lambda\over 8\pi G},\label{eq0020}\\
&&T^r_r=T^t_t-\left[(1+2\beta^{-2} X\right){\eta^2 f'^2\over B^2},\\
&&T^\theta_\theta=T^t_t-\left(1+2\beta^{-2} X\right)
{\eta^2 f^2\over r^2}=T^\phi_\phi,
\end{eqnarray}
and the equation of motion for scalar field 
\begin{eqnarray}
{1\over ABr^2}\left[\left(1+2\beta^{-2} X\right){Ar^2 f'\over B}\right]'-
\left(1+2\beta^{-2} X\right){2 f\over r^2}-\lambda\eta^2f(f^2-1)=0.
\end{eqnarray}

Now we consider exterior solution which makes the energy-momentum tensor components
\begin{eqnarray}
&&T^t_t=T^r_r={\eta^2\over r^2}+{\beta^{-2}\eta^4\over r^4}+{\Lambda\over 8\pi G},\\
&&T^\theta_\theta=T^\phi_\phi=-{\beta^{-2}\eta^4\over r^4}+{\Lambda\over 8\pi G}.
\end{eqnarray}
Because $T^t_t=T^r_r$, we also get $B=A^{-1}$. From the Einstein equation $R^t_t-R/2=8\pi G T^t_t$ we get 
\begin{equation}
(rB^{-2})'=1-8\pi Gr^2 \left({\eta^2\over r^2}+{\beta^{-2}\eta^4\over r^4}\right)-\Lambda r^2,
\end{equation}
and after integration
\begin{equation}
B^{-2}=1-\Delta-{2GM\over r}+{8\pi G\beta^{-2}\eta^4\over r^2} -\frac{\Lambda}{3}r^2,
\label{eq002}
\end{equation}
This solution is valid outside the monopole core, $r>\delta$, with the core given by (with the same method as the previous section),
\begin{equation}
\delta=\frac{2}{\lambda\eta^2}+\sqrt{\left(\frac{2}{\lambda\eta^2}\right)^2+\frac{4}{\beta^2\lambda}}.
\end{equation} 
Here, unlike the previous case, the effect of smaller $\beta^2$ renders the monopole to have a thicker size.

By rescaling with the same method as the previous section, the metric can also be cast into a Reissner-Nordstrom-like metric with global monopole,
\begin{eqnarray}
\label{metricX}
ds^2&=&\left(1-{2GM\over r}+{8\pi G\beta^{-2}\eta^4\over r^2}-{\Lambda\over 3}r^2\right)dt^2-\left(1-{2GM\over r}+{8\pi G\beta^{-2}\eta^4\over r^2}-{\Lambda\over 3}r^2\right)^{-1}dr^2\nonumber\\
&&+\left(1-\Delta\right)r^2 d\Omega^2_2,
\end{eqnarray}
Let us consider the case for $\Lambda=0$. The metric has two roots given by
\begin{equation}
\label{rootsX}
r_\pm=GM\left[1\pm \sqrt{1-\frac{8\pi\eta^4}{M^2 G\beta^2}}\right].
\end{equation}
There is a minimum value of $\beta$ allowed. The real roots exist only when 
\begin{equation}
\label{betamin}
\beta^2>\beta_{crit}^2\equiv{8\pi\eta^4\over GM^2}.
\end{equation}
Below this value, our solution suffers from naked singularity. For $\eta\ll m_P$, where $m_P$ is the Planck mass, this lower bound permits a fairly broad range of small and large $\beta$. The black hole condition~\eqref{bhcon} requires
\begin{equation}
\beta\gtrsim{2\over G^2 M^2}.
\end{equation}
The strongly-coupled regime ($\beta^2<1$) can then produce black hole configuration when $M\gtrsim m_P$.

The roots~\eqref{rootsX} become Reissner-Nordstrom-like horizons when the monopole is encapsulated inside the inner horizon, $\delta<r_-$. This condition reduces to that of~\eqref{betamin}. Thus no additional constraint is imposed. Here the system behaves precisely like the Reissner-Nordstrom black hole. The inner and outer horizons can coalesce and renders the black hole to be extremal. This happens when $\eta=\left({M^2G\beta^2\over 8\pi}\right)^{1/4}$.  

When $\Lambda\neq 0$ there will exist an additional horizon, the cosmological horizon. As in the case of DBI monopole, we avoid solving the quartic equation explicitly. But this cosmological horizon can be seen by considering the case of almost-purely global monopole-de sitter and solving the roots of
\begin{equation}
1+{8\pi G\eta^4\over \beta^2 r^2}-{\Lambda\over 3}r^2\simeq0.
\end{equation}
It differs from the quartic equation\eqref{approxroot} only in the sign of the second term. The roots are
\begin{equation}
r_\pm\simeq\sqrt{3\over 2\Lambda}\sqrt{1\pm\sqrt{1+32\pi G\eta^4\Lambda\over 3\beta^2}}.
\end{equation}
For $\Lambda>0$ the inner horizon that does not exist; inside the cosmological horizon we are exposed to naked singularity. For $\Lambda<0$ the cosmological horizon vanishes but the inner one remains.  

\section{Compactification solutions}

In this section, we study global $k$-monopole solutions but with different metric ansatz. We employ anstaz that enables spontaneous compactification,
\begin{equation}
ds^2=A(r)^2 dt^2-B(r)^2 dr^2-C^2\left(d\theta^2+\sin^2\theta\phi^2\right),\label{eq007}
\end{equation}
with $C$ a constant, and we consider a nonzero cosmological constant. This ansatz is analogous to the case of global string compactification discussed by Gregory\cite{Gregory:1996dd}, albeit in different context. The resulting metrics described below are the lower-dimensional  (and noncanonical) version of global defect compactification discussed in Ref.~\cite{Olasagasti:2001hm}.

\subsection{Compactification by DBI global monopole}

From the Lagrangian and the metric~\eqref{eq007} we get the components of energy-momentum tensor 
\begin{eqnarray}
&&T^t_t=-\beta^2 \left(1-\sqrt{1+{\eta^2\over\beta^2}\left({f'^2\over B^2}+{2f^2\over C^2}\right)}\right)+{\lambda\eta^4\over 4}(f^2-1)^2+{\Lambda\over 8 \pi G}
,\label{eq0010}\\
&&T^r_r=T^t_t-{\eta^2 f'^2/B^2\over\sqrt{1+({\eta^2}/{\beta^2})\left({f'^2}/{B^2}+2{f^2}/{C^2}\right)}},\\
&&T^\theta_\theta=T^t_t-{\eta^2 f^2/r^2\over\sqrt{1+({\eta^2}/{\beta^2})
\left({f'^2}/{B^2}+2{f^2}/{C^2}\right)}}=T^\phi_\phi.
\end{eqnarray}
The components of Ricci tensor becomes
\begin{eqnarray}
&&R^t_t={1\over B^2}\left({A''\over A}-{A'B'\over AB}\right)=R^r_r,~~R^\theta_\theta=-{1\over C^2}=R^\phi_\phi.\label{eq010}
\end{eqnarray}

By setting $f\approx 1$, the components of 
energy-momentum tensor change to 
\begin{eqnarray}
&&T^t_t=T^r_r=-\beta^2\left(1-\sqrt{1+{2\eta^2\over\beta^2 C^2}}\right)+{\Lambda\over\kappa},\\
&&T^\theta_\theta=T^\phi_\phi=T^t_t-{\eta^2/C^2\over \sqrt{1+{2\eta^2\over\beta^2 C^2}}}\label{Tmunucompac},
\end{eqnarray}
with $\kappa\equiv 8\pi G$. To calculate $C^2$, we use $R^t_t-\frac{1}{2}\delta^t_t R = \kappa T^t_t$ 
\begin{equation}
\label{persC}
{1\over C^2}=-\kappa\beta^2\left(1-\sqrt{1+{2\eta^2\over\beta^2 C^2}}\right)+\Lambda.
\end{equation}
Note that in the non-DBI regime, when $\Lambda=0$ we have $\eta=\kappa^{-1/2}$ while the radius $C$ cannot be determined; it is arbitrary. For $\Lambda\neq0$ the radius is given by 
\begin{equation}
C^2={1-\kappa\eta^2\over\Lambda}.
\end{equation}
These results are in agreement with~\cite{Olasagasti:2000gx}.

The general solutions of~\eqref{persC} are
\begin{equation}
\label{Csol}
C^2_{\pm}= {1\over\beta ^2 \kappa  
\left(\eta ^2 \kappa -1\right)+\Lambda
\pm \sqrt{\beta ^2 \kappa ^2 
\left(\beta ^2 \left(\eta ^2 \kappa -1\right)^2+2 \eta ^2 
\Lambda \right)} }.
\end{equation}
Without any loss of generality we can pick $C^2_{+}$. This root must be positive, thus it places constraints on the symmetry-breaking scale $\eta$:
\begin{enumerate}
\item when $\Lambda>0$, 
\begin{equation}
\eta>0,
\end{equation}
\item when $\Lambda=0$, 
\begin{equation}
\eta >\eta_{crit_1}\equiv\sqrt{1/\kappa},\label{critcompac}
\end{equation}
\item when $\Lambda<0$, 
\begin{equation}
\eta\geq\eta_{crit_2}\equiv\sqrt{\frac{\beta ^2 \kappa -\Lambda }{\beta ^2 \kappa ^2}+\sqrt{\frac{\Lambda ^2-2 \beta ^2 \kappa  \Lambda }{\beta ^4 \kappa ^4}}}.
\end{equation}
\end{enumerate}
Note that we can have compactification solution with definite radius even in the absence of $\Lambda$, Eq.~\eqref{critcompac}. This opens up the fate of the super-critical ($\Lambda=0$) DBI global monopole, left undiscussed in~\cite{Liu:2009eh}. For $\eta<\eta_{crit}$ the spacetime develops a $\Delta$-wedge. When $\eta>\eta_{crit}$ the spacetime might spontaneously compactify, $M_4\rightarrow {\mathcal Z}_2\times S^2$. We shall see later that the only possible solution for ${\mathcal Z}$ is the de Sitter space. This phenomenon is reminiscent of the compactification by (super-)massive cosmic strings~\cite{Blanco-Pillado:2013axa, Ortiz:1990tn}, albeit with different co-dimensions. This result is genuinely due to the nonlinearity of our theory, which otherwise non-existent in the ordinary global monopole (for example, see~\cite{Olasagasti:2000gx}).

Now we wish calculate $A$ and $B$ through $R^\theta_\theta-\frac{1}{2}\delta^\theta_\theta R = \kappa T^\theta_\theta$, 
\begin{equation}
{1\over B^2}\left[-{A''\over A}+{A'B'\over AB}\right]=\pm\omega^2,
\end{equation}
where $\pm\omega^2\equiv\kappa T^\theta_\theta$ is now just a constant. Keeping $\omega$ real, the plus and minus sign corresponds to $T^\theta_\theta >0$ and $T^\theta_\theta <0$ respectively. Judging from the Einstein's equations, there is still a freedom in the metric~\eqref{eq007}. We can restrict it by fixing ansatz for $B$. 

The first ansatz we choose is $B=1$, which makes 
\begin{eqnarray}
A(r)=
\begin{cases}
A_0 \sin (\omega r),& \text{if~} T^\theta_\theta >0,\\
A_0 \sinh (\omega r),& \text{if~} T^\theta_\theta <0,
\end{cases}
\end{eqnarray}
and if we define $\chi\equiv \omega r$ and $A_0\equiv1/\omega$ we get 
\begin{eqnarray}
ds^2=
\begin{cases}
\frac{1}{\omega^2}(\sin^2 \chi ~dt^2 - d\chi^2) -C^2 d\Omega_2^2, & \text{if~} T^\theta_\theta >0,\\
\frac{1}{\omega^2}(\sinh^2 \chi ~dt^2 - d\chi^2) -C^2 d\Omega_2^2, & \text{if~} T^\theta_\theta <0.
\end{cases}
\label{eq:0001}
\end{eqnarray}
These are the $dS_2\times S^2$ (Nariai)~\cite{Nariai} and $AdS_2\times S^2$ (Bertotti-Robinson)~\cite{Bertotti:1959pf, Robinson:1959ev} solutions, respectively.

For $T^\theta_\theta=0$, the general solution is $A=C_1 r+ C_2$,
\begin{equation}
ds^2=(C_1 r+ C_2)^2 dt^2 -dr^2 -C^2 d\Omega_2^2.
\label{eq:0002}
\end{equation}
Since this solution has no singularity at $r=0, -C_2/C_1$, and $n\pi/\omega$ with $n=0,1,2,...$, thus we have freedom on $C_1$ and $C_2$. We can set $C_1=0$ to yield (after rescaling $t\rightarrow C_2 t$)
\begin{equation}
ds^2=dt^2 -dr^2 -C^2 d\Omega_2^2,
\end{equation}
The Pleba\'{n}ski-Hacyan ($M_2\times S^2$) metric\footnote{Another possibility is setting $C_2=0$, which yields (after rescaling $t\rightarrow C_1 t$)
\begin{equation}
ds^2=r^2dt^2-dr^2-C^2d\Omega^2_2.
\end{equation}
The two-dimensional space of constant solid angle is a Rindler space. Since we know that Rindler space is flat, then it is equivalent to Minkowski solution.}~\cite{Plebanski}.

The other ansatz of $B$ we can choose is $B=A^{-1}$. This yields
\begin{eqnarray}
A^2=
\begin{cases}
C_2+C_1 r -\omega^2 r^2, & \text{if~} T^\theta_\theta >0,\\
C_2+C_1 r +\omega^2 r^2, & \text{if~} T^\theta_\theta <0,
\end{cases}
\end{eqnarray}
and
\begin{equation}
A^2=C_2+C_1 r, ~~\text{if~} T^\theta_\theta =0,
\end{equation}
with $C_1$ and $C_2$ constants of integration.
Without loss of generality we can set $C_1=0$ and $C_2=1$, 
\begin{eqnarray}
ds^2=
\begin{cases}
(1 -\omega^2 r^2) dt^2 -\frac{dr^2}{(1 -\omega^2 r^2)} - C^2 d\Omega_2^2, & \text{if~} T^\theta_\theta >0,\\
dt^2 -dr^2 - C^2 d\Omega_2^2, & \text{if~} T^\theta_\theta =0,\\
(1 +\omega^2 r^2) dt^2 -\frac{dr^2}{(1 +\omega^2 r^2)} - C^2 d\Omega_2^2, & \text{if~} T^\theta_\theta <0,
\end{cases}
\label{eq:0003}
\end{eqnarray}
which are also Nariai, Pleba\'{n}ski-Hacyan, and Bertotti-Robinson compactifications written in static coordinates, respectively.

\begin{table}[h]
\caption{Conditions for the existence of $T^{\theta}_{\theta}$ as a function of $\eta$ in DBI monopole compactification.}
\begin{tabular}{c||c|c|c}
\hline \rule[-2ex]{0pt}{5.5ex}  & $T^\theta_\theta>0$ & $T^\theta_\theta=0$ & $T^\theta_\theta<0$ \\ 
\hline \hline \rule[-2ex]{0pt}{5.5ex} $\Lambda>0$\ \ \; & $\eta>0\ \ $ &\ \  does not exist\ \ \ \ &\ \  does not exist\ \  \ \ \\ 
\hline \rule[-2ex]{0pt}{5.5ex} $\Lambda=0$\ \ \; & $\eta> \eta_{crit_1}\ \ $ & $\eta \leq \eta_{crit_1}\ \ \ \ $ &\ \  does not exist\ \ \ \ \\ 
\hline \rule[-2ex]{0pt}{5.5ex} $\Lambda<0$\ \ \; & \;$\eta >\eta_{crit_2}\ \ $\; & \; $\eta =\eta_{crit_2}\ \ \ \ $ \; & \;$\eta <\eta_{crit_2}$\ \ \ \ \; \\ 
\hline 
\end{tabular} 
\label{table:3}
\end{table}
Notice that the resulting solutions are all product of two $2$-dimensional spaces of constant curvatures. In this sense, we have produced spontaneous compactification solutions of $4d$ global DBI monopole,
\begin{equation}
{\mathcal Y}_4\rightarrow{\mathcal Z}_2\times S^2,
\end{equation}
where ${\mathcal Y}$ and ${\mathcal Z}$ can each be two-dimensional de Sitter, Minkowski, or Anti-de Sitter spaces. As suggested in~\cite{Blanco-Pillado:2013axa} these are the four-dimensional analogue of the flux compactification discussed, for example, in~\cite{RandjbarDaemi:1982hi, Salam:1984cj}. The $4d$ space has a cosmological constant $\Lambda$, while the $2d$ space gains the effective corresponding constant given by $\omega$. To check which compactification channels are actually present in this theory, we should investigate the constraints of $\omega$'s existence in relation with $\Lambda$. Here, we substitute the radius solution~\eqref{Csol} into $T^{\theta}_{\theta}$,~\eqref{Tmunucompac}. Solving the polynomial equations, we end up with the range of $\eta$ allowed that constraints the existence of two-dimensional vacua of constant curvature, as shown in Table~\eqref{table:3}. 
Combined with the conditions for $C^2>0$, we can see that not all possible compactifications above exist. Take the $M_4\rightarrow M_2\times S^2$ channel, for example. The vanishing of $\omega$ requires $\eta\leq\eta_{crit}$, as shown in Table~\eqref{table:3}. However, this contradicts~\eqref{critcompac} which forces $\eta>\eta_{crit}$. Thus we conclude that such channel does not exist. The possible compactification channels are listed below
\begin{eqnarray}
dS_4 &\longrightarrow& dS_2 \times S^2,\label{m4m1}\\
M_4 &\longrightarrow& dS_2 \times S^2,\label{m4m2}\\ 
AdS_4 &\longrightarrow&
\begin{cases}
dS_2\times S^2,\\
M_2\times S^2.\label{m4m3}
\end{cases}
\end{eqnarray}
Note that Eq.~\eqref{m4m2} provides a possible fate of spacetime around super-critical global DBI monopole. Here we can conclude that beyond $\eta>\eta_{crit}$ static solution ceases to exist. Instead we have an inflating solution along the two-dimensional spacetime ($dS_2$) and spherical compactification ($S^2$) along the other two spatial dimensions. 
\subsection{Compactification by power-law global monopole}
Now, we consider compactification with metric (\ref{eq007}) and kinetic term (\ref{eq008}) which components of energy-momentum tensor are
\begin{eqnarray}
&&T^t_t=X+\beta^{-2} X^2+{\lambda\eta^4\over 4}(f^2-1)^2+{\Lambda\over\kappa},\\
&&T^r_r=T^t_t-(1+2\beta^{-2} X){\eta^2f'^2\over B^2},\\
&&T^r_r=T^t_t-(1+2\beta^{-2} X){\eta^2f^2\over C^2}.
\end{eqnarray}
The components of Ricci tensors are the same as (\ref{eq010}). The exterior condition reduces them into
\begin{eqnarray}
&&T^t_t=T^r_r={\eta^2\over C^2}+{\beta^{-2}\eta^4\over C^4}+{\Lambda\over\kappa},\\
&&T^\theta_\theta=T^\phi_\phi=-{\beta^{-2}\eta^4\over C^4}+{\Lambda\over\kappa}.
\end{eqnarray}

From the same method as before we get 
\begin{eqnarray}
{1\over C^2}={\kappa\eta^2\over C^2}+{\kappa\beta^{-2}\eta^4\over C^4}+\Lambda.
\end{eqnarray}
For $\Lambda=0$ we get 
\begin{eqnarray}
\label{csol0}
C^2={\beta^{-2}\kappa\eta^4\over 1-\kappa\eta^2},
\end{eqnarray}
which puts constraint $\eta>\eta_{crit_1}\equiv\sqrt{1/\kappa}$ in order to have a sensible compactification.
For $\Lambda\neq 0$ we get 
\begin{eqnarray}
\label{csoltak0}
C^2_\pm=\frac{
(1-\kappa\eta^2)\pm\sqrt{
(\kappa\eta^2-1)^2-4\Lambda\kappa\eta^4\beta^{-2}
}}{2\Lambda}.
\end{eqnarray}
After choosing $C^2_+$ its positivity condition requires:
\begin{enumerate}
\item for $\Lambda>0$, 
\begin{equation}
\eta\leq {1\over\sqrt{\kappa+2\sqrt{\beta^{-2}\kappa|\Lambda|}}}\equiv \eta_{crit_3},
\end{equation}
\item for $\Lambda<0$, 
\begin{equation}
\eta>0.
\end{equation}
\end{enumerate}

\begin{table}[h]
\caption{Conditions for the existence of $T^{\theta}_{\theta}$ as a function of $\eta$ in power-law monopole compactification.}
\begin{tabular}{c||c|c|c}
\hline \rule[-2ex]{0pt}{5.5ex}  & $T^\theta_\theta>0$ & $T^\theta_\theta=0$ & $T^\theta_\theta<0$ \\ 
\hline \hline \rule[-2ex]{0pt}{5.5ex} $\Lambda>0$\ \ \; & \; $\eta<\eta_{crit_3}\ \ $ \; & \; $\eta=\eta_{crit_3}\ \ $ \; & \; \ \ $\eta>\eta_{crit_3}$\ \ \; \\ 
\hline \rule[-2ex]{0pt}{5.5ex} $\Lambda=0$\ \ \; & \; \ \ does not exist\ \ \ \ \; & \; $\eta = 1/\kappa^{1/4}$ \; & \; \ \ \ $\eta \neq 1/\kappa^{1/4}$\ \  \; \\ 
\hline \rule[-2ex]{0pt}{5.5ex} $\Lambda<0$\ \ \; & \; \ \ does not exist\ \ \ \  \; &\; \ \ does not exist\ \ \ \ \; & \; $\eta>0$\ \  \; \\ 
\hline 
\end{tabular} 
\label{table:2}
\end{table}
We can use the same method as in the previous section in calculating the metric solutions which will arrive at the same results ((\ref{eq:0001}), (\ref{eq:0002}), and (\ref{eq:0003})), thus we can continue to finding the conditions for symmetry-breaking scale. By inserting the radius solutions~\eqref{csol0} and \eqref{csoltak0} into $T^\theta_\theta$ we can solve the polynomial equations to extract the range of $\eta$ allowed that constraints the existence of compactification solutions. The results are shown in Table \ref{table:2}. This result is then combined with the conditions for $C^2>0$ to see which compactification channels are theoretically possible. Note that for $\Lambda=0$ the criteria for $\omega^2=0$ does not contradict the constraint given from the radius since $1/\kappa^{1/4}>1/\sqrt{\kappa}$. Thus,  we can list the possible compactification channels in this theory as
\begin{eqnarray}
AdS_4 &\longrightarrow& AdS_2 \times S^2,\label{p1}\\
\nonumber\\
M_4 &\longrightarrow& 
\begin{cases}
M_2 \times S^2,\\
AdS_2\times S^2,
\end{cases} 
\end{eqnarray}
\begin{eqnarray} 
dS_4 &\longrightarrow&
\begin{cases}
dS_2\times S^2,\\
M_2\times S^2.\label{p3}
\end{cases}
\end{eqnarray}
Here the flat super-critical global monopole is possible to compactify the spacetime into an $M_2\times S^2$. Since this is a static spacetime, we conjecture that this channel is unstable.

\section{Conclusions}

In this paper we have established analytical solutions of global $k$-monopoles with cosmological constant for various spacetime topology. We specifically consider two types of $k$-monopole: the DBI and the power-law types. These are monopoles with noncanonical kinetic terms. It is proposed~\cite{Babichev:2006cy, Babichev:2007tn} that these kind of defects might have been formed in the very early universe, and that due to their non-standard form their existence might not be ruled out by the present observations. For each, we study the static conical topology and compactification solutions.

For the static case, our results are the analytical and the asymptotically dS/AdS version of the otherwise flat numerical solutions studied in~\cite{Jin:2007fz, Liu:2009eh}. We found that these $k$-monopoles produce conical spacetime in their surrounding with deficit angle given by $\Delta=8\pi G\eta^2$, independent of the nonlinear coupling constant $\beta$. Nonlinearity of the global monopole does not affect the topological property of the surrounding spacetime. We next analyzed the resulting event and cosmological horizons formed when a black hole swallows a global $k$-monopole. In order for this to happen, we require $M\gtrsim m_P$. For DBI monopole, it is shown that the inner root is actually complex. Thus it behaves more like a Schwarzschild black hole. The DBI-monopole black hole behaves just like the ordinary BV black hole. In the case of power-law monopole, on the other hand, we found that both roots are real. Here the monopole behaves like a Reissner-Nordstrom black hole, only that it is scalarly, not electromagnetically, charged. When the cosmological constant is turned on, we have an additional horizon for both cases. Assuming $\Lambda\gg M^{-2}$, when $\Lambda>0$ the spacetime around a DBI-monopole has two horizons (the monopole and the cosmological horizons) which coincide when $\Lambda={3\beta^2\over16G\pi\eta^4}$, while for a power-law monopole inner horizon does not exist so that inside the cosmological horizon one is exposed to naked singularity.  For $\Lambda<0$, both the DBI and power-law monopoles lose their cosmological horizon which renders them Schawarzschild-like.

The next question is: what happens when $\eta>\eta_{crit}$? It is argued in Refs.~\cite{Cho:2003gn, Liebling:1999bb} that in flat spacetime super-critical global monopole will develop curvature singularity and therefore no static solution exists. To cure singularity it is proposed that the monopole is allowed to inflate~\cite{Cho:2003gn, Cho:1997rb}, which eventually results in the topological inflation~\cite{Vilenkin:1994pv, Linde:1994hy}. We tried to answer the same question for $k$-monopole, and our investigation reveals that super-critical global $k$-monopole is able to compactify its surrounding spacetime into a product of two two-dimensional maximally symmetric spaces. This is an example of lower-dimensional spontaneous compactification, for example as discussed in~\cite{Blanco-Pillado:2013axa, BlancoPillado:2010uw}, where here the $2$-sphere is threaded against collapse by the flux coming from the scalar field. This singularity-free spacetime is interpreted as non-static solutions. The effective two-dimensional cosmological constant is provided by $\omega^2$. For the DBI monopole we have $M_4\rightarrow dS_2\times S^2$, a Nariai compactification, while for the power-law monopole we can have $M_4\rightarrow M_2\times S^2$ (Pleba\'{n}ski-Hacyan compactification) and $M_4\rightarrow AdS_2\times S^2$ (Bertotti-Robinson compactification). Since no static solution can exist for super-critical monopole, we conjecture that the Pleba\'{n}ski-Hacyan solution is unstable. Here we only focus on the compactification with positive constant curvature. The possibility of hyperbolic space (negative constant curvature), for example the anti-Nariai space ($AdS_2\times H^2$), will be left for future work.

For $\Lambda\neq 0$, we can have a set of possible compactification channels as shown in Tables~\eqref{table:3} and \eqref{table:2}. Note that not all of these channels exist. When taking into account constraints from the compact radius, the allowed channels for DBI monopole are listed in Eqs.~\eqref{m4m1}-\eqref{m4m3}. For power-law monopole, we have the allowed channels are given in Eqs.~\eqref{p1}-\eqref{p3}. Note that these analytical solutions are all subject to the vacuum approximation ($f\approx 1$), which is not an exact condition. It remains to be clarified by numerical analysis that such compactification solutions really exist. It may be that they develop topological inflation before starting to compactify.

One thing left unaddressed in this paper is the question of stability. As we conjectured above, the channel $M_4\rightarrow M_2\times S^2$ might be unstable\footnote{However, see Ref.~\cite{Blanco-Pillado:2013axa}.}. It is also unclear whether all these compactification channels are actually stable. It is possible that the two-dimensional de Sitter space produced are the local maximum of the effective radion potential. Thus, it is unstable against transdimensional tunneling ~\cite{BlancoPillado:2009di, BlancoPillado:2009mi}. The difficulty in investigating stability is precisely because in two dimensions the action cannot be written in the Einstein frame, so that we cannot extract out the radion potential, as in the case of $dS_6\rightarrow dS_2\times S^4$ decay studied in Ref.~~\cite{BlancoPillado:2009mi}. We shall work on it in the forthcoming publication.

\section{Acknowledgements}

We thank Ardian Atmaja and Jose Blanco-Pillado for useful discussions and comments on the early manuscript. This work was partially supported by the Research-Cluster-Grant-Program of the University of Indonesia No.~1862/UN.R12/HKP.05.00/2015.


\end{document}


\title{Erratum for ``Gravity of a noncanonical global monopole: conical topology and compactification"}

\author{Ilham Prasetyo}
\email{ilham.prasetyo51@ui.ac.id}
\author{Handhika S. Ramadhan}
\email{hramad@ui.ac.id}
\affiliation{Departemen Fisika, FMIPA, Universitas Indonesia, Depok 16424, Indonesia. }

\def\changenote#1{\footnote{\bf #1}}

\begin{abstract}

We revisit our calculations in~\cite{Prasetyo:2015bga} and found that there are some mistakes in the solutions of power-law global monopole black holes and compactification channels. We hereby present the following erratum correcting our error and modifying our conclusions.

\end{abstract}

\maketitle
\thispagestyle{empty}
\setcounter{page}{1}

\section{Introduction}


In our recent paper~\cite{Prasetyo:2015bga} (and in~\cite{Prasetyo:2016nng}) we studied analytically the gravity of noncanonical global monopole, a nonlilnear generalization of Barriola-Vilenkin solution~ \cite{Barriola:1989hx}. Specifically, we investigate the black hole and compactification solutions of two types of $k$-global monopole~\cite{Babichev:2006cy}: the Dirac-Born-Infeld (DBI) and the power-law types. While such gravitational solutions have been studied numerically in~\cite{Jin:2007fz, Liu:2009eh}, our work serves as a complimentary result that provides explanation for the black hole horizons and compactification channels left undiscussed there.

It is later realized that our results for the power-law global monopole contains error, both the mathematical solutions and the conclusion. The result for DBI global-monopole, on the other hand, are free from error as far as we have checked. This short letter is therefore dedicated to the erratum, and we hope this will clarify the confusion that might be caused by our previous result.

\section{Black hole solutions}

In Eq.(52) of~\cite{Prasetyo:2015bga} we have
\begin{equation}
r_\pm\simeq\sqrt{3\over 2\Lambda}\sqrt{1\pm\sqrt{1+{32\pi G\eta^4\Lambda\over 3\beta^2}}}.
\end{equation}
For $\Lambda<0$ we stated that the cosmological horizon vanishes but the inner one remains. This is inaccurate, as can easily be shown. The almost-purely-global-monople-AdS black hole suffers from naked singularity.

\section{Compactification solutions}

For the compactification, our Eqs.(87) and (88) in~\cite{Prasetyo:2015bga} is wrong. the correct results should be as follows
\begin{equation}
\eta^2 \leq \frac{1}{8} \left[\frac{2 \beta ^2 }{2 \pi  \beta ^2 G-\Lambda }+\sqrt{\frac{2\beta ^2 \Lambda }{\pi G \left(\Lambda -2 \pi  \beta ^2 G\right)^2}}\right]
\equiv \eta^2_{c+}
\label{eq:condk2-0}
\end{equation}
for $0<G<\frac{\Lambda }{2 \pi  \beta ^2}$, 
\begin{equation}
\eta^2 \leq \frac{1}{16 \pi  G}\equiv\eta_{crit}^2/2 \label{eq:condk2-1}
\end{equation}
for $G=\frac{\Lambda }{2 \pi  \beta ^2}$, and 
\begin{equation}
\eta^2 \leq \frac{1}{8} \left(\frac{2 \beta ^2 }{2 \pi  \beta ^2 G-\Lambda }- \sqrt{\frac{2\beta ^2  \Lambda }{\pi G \left(\Lambda -2 \pi  \beta ^2 G\right)^2}}\right)
\equiv \eta^2_{c-}
\label{eq:condk2-2}
\end{equation}
for $G>\frac{\Lambda }{2 \pi  \beta ^2}$. 

The Table II in~\cite{Prasetyo:2015bga} is also inaccurate. Our recent calculation is shown in Table \ref{table:tangk}
\begin{table}[h]
\caption{Conditions for the existence of $T^{\theta}_{\theta}$ as a function of $\eta$ in power-law monopole compactification.}
\begin{tabular}{cccc}
		\hline \rule[-2ex]{0pt}{5.5ex}  & \,\,$dS_2\times S^2$\,\, & \,\,$M_2\times S^2$\,\, & \,\,$AdS_2\times S^2$ \\ 
		\hline \hline \rule[-2ex]{0pt}{5.5ex} $\Lambda>2 \pi  \beta^2  G$\ \ \, & $\eta^2 < \eta^2_{c+}$ \ \  &\ \  $\eta^2 = \eta^2_{c+}$ \ \ \ \ &\ \  Can not happen \ \  \ \ \\ 
		\hline \rule[-2ex]{0pt}{5.5ex} $\Lambda=2 \pi  \beta^2  G$\ \ \, & $\eta^2<\frac{1}{16 \pi  G}$ \ \  &\ \  $\eta^2=\frac{1}{16 \pi  G}$ \ \ \ \ &\ \  Can not happen \ \  \ \ \\ 
		\hline \rule[-2ex]{0pt}{5.5ex} $\Lambda<2 \pi  \beta^2  G$\ \ \, & $\eta^2 < \eta^2_{c-}$ \ \  &\ \  $\eta^2 = \eta^2_{c-}$ \ \ \ \ &\ \  Can not happen  \ \  \ \ \\
		\hline \rule[-2ex]{0pt}{5.5ex} $\Lambda=0$\ \ \, & Can not happen & cannot happen \ \ \ \  &\ \  $\eta^2 < \eta^2_{crit}$ \ \ \ \ \\ 
		\hline \rule[-2ex]{0pt}{5.5ex} $\Lambda<0$\ \ \, & \, {} \, & \, Can not happen \ \ \ \  \, & \, {} \ \ \ \ \, \\ 
		\hline 
\end{tabular}
\label{table:tangk}
\end{table}
Combining these with conditions~\eqref{eq:condk2-0}-\eqref{eq:condk2-2}, the possible channels are
\begin{eqnarray}
dS_4 &\longrightarrow&
\begin{cases}
dS_2\times S^{2},\\
M_2\times S^{2},
\end{cases}\\
M_4 &\longrightarrow& AdS_2\times S^{2}.
\end{eqnarray}
\section{Conclusions}

This erratum serves as a supplementary article for reading~\cite{Prasetyo:2015bga, Prasetyo:2016nng}, where some errors in the global monopole with power-law kinetic terms were spotted. The conclusions are two-fold. First, we clarify that the black hole with almost-purely  global monopole ($\Lambda\gg M^{-2}$) suffers from naked singularity. Second, our compactification channels are devoid of unstable Pleba\'{n}ski-Hacyan compactification from $4D$ Minkowski, in contrast to what was reported earlier. One implication we would like to draw is that our previous conclusion which said that the global $k$-monopole can compactify its surrounding spacetime when it is super-critical should be corrected. A thorough calculation shows that the compactification can happen before the monopole reaches its super-critical stage. As shown in Table~\ref{table:tangk} the $4d$ flat monopole can compactify if $\eta^2 < \eta^2_{crit}$, not $\eta\neq{1\over\eta^{1/4}}$. 

\section{Acknowledgements}

We thank Anto Sulaksono and Terry Mart for the discussion and useful suggestions.
